\documentclass[journal=apchd5,manuscript=article]{achemso}

\usepackage{chemformula} 
\usepackage[T1]{fontenc} 
\usepackage{overpic}
\usepackage[normalem]{ulem}
\usepackage{hyperref}
\hypersetup{colorlinks=true,allcolors=black}

\author{Pedro Ninhos}
\affiliation[University of Southern Denmark]
{POLIMA---Center for Polariton-driven Light--Matter Interactions, University of Southern Denmark, Campusvej 55, DK-5230 Odense M, Denmark}
\author{Christos Tserkezis}
\affiliation[University of Southern Denmark]
{POLIMA---Center for Polariton-driven Light--Matter Interactions, University of Southern Denmark, Campusvej 55, DK-5230 Odense M, Denmark}
\author{N.~Asger~Mortensen}
\affiliation[University of Southern Denmark]
{POLIMA---Center for Polariton-driven Light--Matter Interactions, University of Southern Denmark, Campusvej 55, DK-5230 Odense M, Denmark}
\alsoaffiliation[D-IAS]
{D-IAS---Danish Institute for Advanced Study, University of Southern Denmark, Campusvej 55, DK-5230 Odense M, Denmark}
\author{Nuno~M.~R.~Peres}
\affiliation[University of Southern Denmark]
{POLIMA---Center for Polariton-driven Light--Matter Interactions, University of Southern Denmark, Campusvej 55, DK-5230 Odense M, Denmark}
\alsoaffiliation[Universidade do Minho]
{Centro de F\'{\i}sica (CF-UM-UP) and Departamento de F\'{\i}sica, Universidade do Minho, P-4710-057 Braga, Portugal}
\alsoaffiliation[INL]
{International Iberian Nanotechnology Laboratory (INL), Av Mestre Jos\'e Veiga, 4715-330 Braga, Portugal}
\email{peres@fisica.uminho.pt}

\title[Tunable exciton polaritons]
  {Tunable Exciton Polaritons in Band-Gap Engineered Hexagonal Boron Nitride}

\keywords{hexagonal boron nitride, super lattice, sub-gap optical response, exciton polaritons, photonics}

\newcommand*{\citen}[1]{%
  \begingroup
    \romannumeral-`\x 
    \setcitestyle{numbers}%
    \cite{#1}%
  \endgroup   
}

\begin{document}

\begin{tocentry}

\centering
\includegraphics{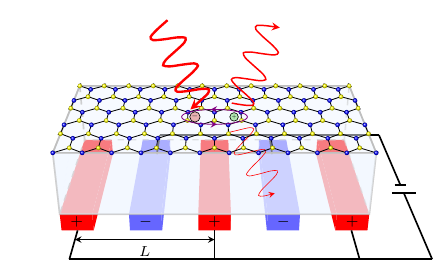}

\end{tocentry}

\begin{abstract}
We show that hexagonal boron nitride (hBN), a two-dimensional insulator,
when subjected to an external superlattice potential forms a paradigm
for electrostatically tunable excitons in the near- and mid-ultraviolet (UV).
With a combination of analytical and numerical methods, we
see that the imposed potential has three consequences: (i) it renormalizes the effective
mass tensor, leading to anisotropic effective masses; (ii) it renormalizes
the band gap, eventually reducing it; (iii) it reduces the exciton binding
energies. All these consequences depend on a single dimensionless parameter,
which includes the product of strength of the external potential with its 
period. In addition to the excitonic energy levels, we compute the optical
conductivity along two orthogonal directions, and from it the absorption 
spectrum. The results for the latter show that our system is able to mimic
a grid polarizer. These characteristics make one-dimensional hBN
superlattices a viable and meaningful platform for fine-tuned polaritonics
in the UV to visible spectral range.
\end{abstract}

\section{}

The properties of two-dimensional (2D) materials are characterized by
a richness of physical phenomena, encompassing unusual elastic~\cite{Jiang_2020}, electronic~\cite{Dutta_2023}, and optical
properties~\cite{Jia_2019}. An exemplary case of enhanced light--matter
interactions in 2D materials is provided by graphene~\cite{Gan_2012},
whose transmission in the visible range of the spectrum is characterized
by the value of the fine structure constant alone, a peculiar case in
materials. Concerning the electronic properties in 2D materials, we can
find semi-metals~\cite{Novoselov_2007}, conductors~\cite{Miller_2023},
semiconductors and insulators~\cite{Zeng_2013,Huang_2022}, strange
metals~\cite{Cao_2020}, metallic ferromagnets~\cite{Jiang_2018,Yu_2022},
and superconductors~\cite{Qiu_2021,Jarillo_2018,Jarillo_2021,Cao_2018}.
Therefore, 2D materials have become a condensed-matter physicist's
playground due to this wealth of physical properties. Some systems,
such as hexagonal boron nitride (hBN), have an interesting story: the
first application of hBN was as an encapsulating medium for graphene
and other 2D materials, due to its ultra-flat surface and low density of
defects~\cite{Abidi_2019}. Such use allows to increase the mobility of
charge carriers~\cite{Banszerus_2016} and produce well defined (large
quality factor $Q$) electromagnetic (EM) resonances in polaritonic devices.
Later, it became clear that hBN is interesting in its own right when
light--matter interaction is considered~\cite{Rizzo_2023,Elias_2019}.
The interaction with infrared radiation due to polar phonons in hBN
gives rise to large-$Q$ polaritonic resonances used for example in gas
sensing~\cite{Xu_2023}. At the same time, the interaction of ultraviolet
(UV) radiation with the electrons in hBN has promoted the field of UV 
photonics~\cite{Song_Li_2022}.

\begin{figure*}[t]
\centering
\includegraphics[width=1.0\textwidth]{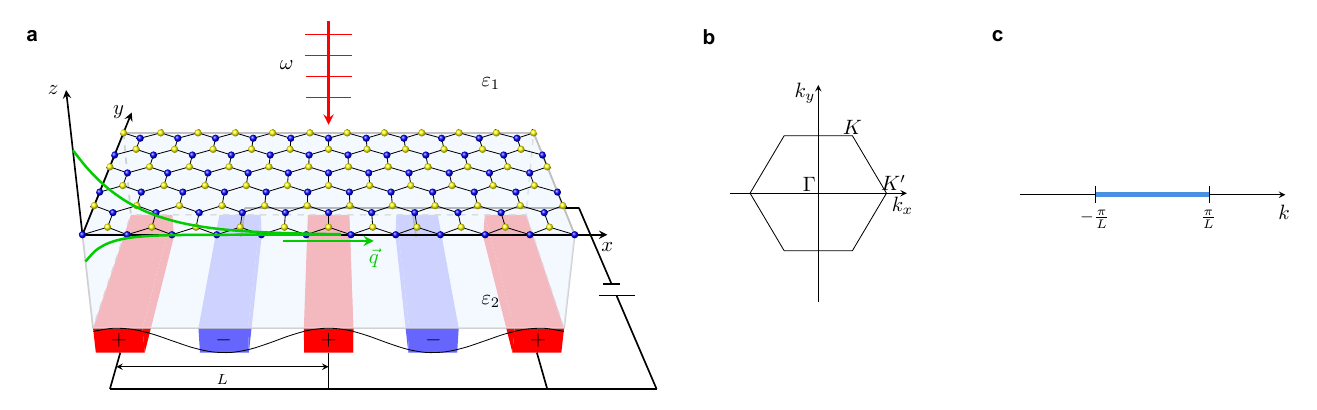}
\caption{(a) An hBN monolayer over a substrate, subjected to an external
periodic potential with period $L$ along the $x$-direction, under a
time-periodic drive of a linearly polarized optical field. The electrostatic gating serves illustrative purposes, as one of the possibilities for achieving the periodic potential. (b) Brillouin zone (BZ) of the original monolayer hBN, and
(c) the BZ of the superlattice resulting from applying the 1D potential
of (a).}

\label{fig:hBN_monolayer}
\end{figure*}

2D materials can be combined by stacking several layers, forming
van~der~Waals heterostructures; materials consisting of few layers
have proven to be a fertile theoretical and experimental template
for the study of unconventional physics~\cite{Geim:2013}. On one hand,
heterostructures are promising for tailoring optoelectronic properties
of 2D insulators (such as hBN)~\cite{Rizzo_2023}. On the other hand, 
homostructures reveal quite interesting features, distinct from those
of the initial isolated constituents. From the example of AB-stacking
graphene bilayer, it was realized that a band gap can be opened and
controlled electrostatically, using a voltage 
bias~\cite{Amorim_2022,Castro_2007}. In twisted bilayer graphene, it
was observed that a superlattice potential has a tremendous impact on
the electronic and optical properties of 2D 
materials~\cite{Bistritzer_2011,Castro_Neto_2007,Kar_2021}. Armed with
this information, it is possible to envision a combination of both
electrostatic and superlattice potentials such that the electronic and
optical properties of 2D materials can be externally controlled. One can
manipulate the electronic and optical properties of a material by
band-gap engineering~\cite{Azadani_2020}. For example, it is possible
to change the gap of bulk hBN by femtosecond laser pulses~\cite{Xue_2021}.
For monolayer hBN, the band gap can be reduced by cladding the hBN between
two graphene layers~\cite{Thygesen_2017}, or by
functionalization~\cite{Das_2012}.

A previous study on nonlocal quantum plasmonic effects
in a graphene one-dimensional (1D) superlattice showed that the external
potential can influence the plasmonic properties of graphene~\cite{Stauber_2020}. Namely, the plasmonic excitations are not
affected by the potential along the direction perpendicular to the
modulation, but those along the direction of the modulation are drastically
modified.
In this context, it is pertinent to ask if one can manipulate the
physical properties of hBN, the 2D material considered in this work, by
forming a superlattice, resulting from combining an hBN lattice with an
electrostatic periodic potential (Figure~\ref{fig:hBN_monolayer}a). As a
pristine 2D lattice, hBN is a wide-band-gap insulator in which boron (B)
and nitrogen (N) atoms are arranged in a honeycomb lattice, bounded by
strong covalent bonds. Unlike graphene, hBN presents a large band gap at
the $\mathrm{K}$ and $\mathrm{K}'$-points of the Brillouin zone (BZ) (see
Figure~\ref{fig:hBN_monolayer}b), and the low-energy dispersion is quadratic
and not linear, as it is in graphene. In a recent work~\cite{Roman_2021},
a direct measurement of the density of states of monolayer hBN by scanning
tunneling microscopy revealed a band gap of $6.8 \pm 0.2$\,eV, while optical
spectroscopy revealed an exciton binding energy of $0.7\pm 0.2$\,eV. With
hBN being a large-band-gap insulator, excitonic effects play an important
role in its optical properties in the deep UV range~\cite{Ferreira_2019}.
This was shown when an experimental measurement of photoluminescence of
monolayer hBN on quartz~\cite{Elias_2019} reported an optical resonance
around 6.1\,eV, that can only be explained by taking excitonic effects
into account. 

When studying the sub-gap optical properties of hBN, we cannot ignore the
fact that the exciton can excite an EM mode confined at the surface of the
2D material, called exciton polariton. The intriguing properties of polaritons
allow for the detection of small changes in the refractive index of the 
surrounding material, rendering the 2D material viable for the fabrication of
optoelectronic sensors. Given the energy scales of the band gap and exciton 
binding energies of hBN, one can think of it as a good candidate for the 
detection of bio-molecules that present resonances in the UV, such as
cyclic $\beta$-helical peptides, relevant for the biophysics of the human
body~\cite{Fears_2013}.

In this work, we propose a means of manipulating the optoelectronic
properties of hBN by applying a 1D periodic potential on a monolayer,
as sketched in Figure~\ref{fig:hBN_monolayer}a. As we will see, the
resultant superlattice has distinct electronic properties from the pristine
hBN monolayer. The layer is deposited on a quartz substrate of permittivity
$\varepsilon_{2}$, and is illuminated from above with monochromatic light,
linearly polarized along either the $x$ or the $y$-axis. This driving field
allows the formation of excitons that, by coupling with light, give rise to
exciton polaritons propagating along the sample.
By externally tuning the applied potential, either through
its periodicity or its strength (e.g. the gating voltage), the emergence of
such polaritons can be engineered over a wide frequency window, from the UV
to the infrared, thus providing an efficient means for making the excitonic
response of hBN relevant to applications in optics and polaritonics.

\section{Results}\label{sec:Results}

\subsection{Renormalized Hamiltonian}

The application of a 1D periodic potential $V(x)$ on top of monolayer hBN
results in the low-energy Hamiltonian
\begin{equation}
\label{eq:Hamiltonian_results}
H = \hbar v_{F}(\sigma_{x}q_{x}+\sigma_{y}q_{y})+M\sigma_{z} + V(x),
\end{equation}
where the terms with the Pauli matrices $\sigma_{x/y}$ times momentum
$q_{x/y}$ correspond to the low-energy Hamiltonian of hBN, with Fermi
velocity $v_F$ given by $\hbar v_F = 5.06$\,eV$\cdot${\AA}, and
$M = 1.96$\,eV being half of the density functional theory (DFT) band gap.
These parameters were determined previously by
Ferreira et al.~ \cite{Ferreira_2019} through first principle calculations,
such as DFT and the $GW$ method. To study excitonic physics, it is, in
general, sufficient to work with an effective low-energy model, justifying
our choice for the starting Hamiltonian of eq~\ref{eq:Hamiltonian_results}.
The potential $V(x)$ can be realized, for example, by an array of gated
metallic rods, very long in length and arranged in parallel, as shown in
Figure~\ref{fig:hBN_monolayer}a.
With the introduction of the external potential, we expect to obtain
several side-bands as a consequence of the resultant emergent superlattice. Since
the Hamiltonian of eq~\ref{eq:Hamiltonian_results} is not written in
the most convenient form, as the potential is a function of real-space
coordinates, while all other terms are given in momentum space, it is
useful to apply a unitary transformation as described in the Methods
section. We further assume a potential of the form $V(x) = V_0 \cos(G_0 x)$,
where $V_0$ is its amplitude or strength, and $G_0 = 2\pi/L$ is the
length of the primitive vector of the reciprocal superlattice and
defines the period of the potential, $L$. Even though 
Figure~\ref{fig:hBN_monolayer}a depicts the potential with a step-wise
form, 
in what follows we assume a cosine form, as it allows
obtaining an analytical expression for the emergent low-energy Hamiltonian;
considering a step-wise potential in our calculations would not make a
significant difference, as the cosine potential can be seen as the first
terms in the Fourier expansion of a step-wise function, and those components
are always the strongest ones.
Only through analytical calculations, we can obtain the renormalized
low-energy Hamiltonian,
\begin{equation}\label{eq:renormalized_Hamiltonian_results}
\bar{H} =
\hbar v_{F}
[\sigma_{x}q_{x}+\sigma_{y}J_{0}(\beta)q_{y}] -
\sigma_{z} M J_{0}(\beta),
\end{equation}
where $\sigma_{z}$ is the third Pauli matrix, $J_0(\beta)$ is the Bessel
function of the first kind of zeroth order, computed at $\beta = V_0 L/\pi
\hbar v_F$, a parameter that only depends on the product between the
strength and period of the potential. In
eq~\ref{eq:renormalized_Hamiltonian_results}, we notice that for a 
quasiparticle propagation along the $y$-direction the group velocity is
renormalized as $v_{F} \rightarrow J_{0}(\beta)v_{F}$. Since $\vert J_{0}
(\beta) \vert < 1$, the slope of the dispersion along the $y$-direction is
smaller than along the $x$-direction. This gives rise to an anisotropic
Dirac cone. Also, the DFT band gap $E_g = 2M$ changes as $E_g \rightarrow
E_g \vert J_{0}(\beta)\vert$. Therefore, it is possible to reduce the
pristine gap of hBN by choosing an appropriate value for $\beta$. With
this mechanism we can control both the gap and the excitonic states,
which now become the excitons of a synthetic anisotropic material, that
resembles naturally occurring anisotropic materials, such as phosphorene.

From the renormalized low-energy Hamiltonian of
eq~\ref{eq:renormalized_Hamiltonian_results}, we can infer the
low-energy dispersion
\begin{equation}
E_{s} \left(\mathbf{q}\right) = 
\frac{s}{2}
\sqrt{E_g^2 J_{0}^{2} (\beta) + 
4 \hbar^{2} v_{F}^{2} 
\left[q_{x}^{2} + q_{y}^{2} J_{0}^{2}(\beta)\right]},
\quad 
\text{with} \quad 
s=+/-,
\end{equation}
which is now anisotropic, while we recover the isotropic case for
$\beta \rightarrow 0$. This deformed paraboloid has two distinct effective
masses $m_x = m^* |J_{0}(\beta)|$ and $m_y = m^*/ |J_{0}(\beta)|$ along
the $x$ and $y$-directions, respectively, where $m^*=E_g/4v_F^2$ is the
electron effective mass in pristine hBN. The eigenproblem
$\bar{H} |s,\mathbf{q} \rangle = E_{s}(\mathbf{q})|s,\mathbf{q} \rangle$
has an analytical solution for the spinors
\begin{subequations}
\label{eq:spinors}
\begin{align}
|+,\mathbf{q}\rangle &= 
\left(\begin{array}{c}
\mathrm{e}^{-i\theta_{q} \left(\beta\right)}
\sin\left(\frac{\xi_{q}}{2}\right) \\ 
\cos \left(\frac{\xi_{q}}{2}\right)
\end{array}\right) , 
\\
|-, \mathbf{q}\rangle & = 
\left(\begin{array}{c}
\cos \left(\frac{\xi_{q}}{2}\right) \\
-\mathrm{e}^{i\theta_{q}\left(\beta\right)}
\sin\left(\frac{\xi_{q}}{2}\right)
\end{array}\right),
\end{align}
\end{subequations}
where $\xi_{q} = \arctan \left(\frac{2\hbar v_{F}q\left(\beta\right)}{J_{0}(\beta)E_{g}}\right)$, with $q\left(\beta\right) = \sqrt{q_{x}^{2}+J_{0}^{2}(\beta)q_{y}^{2}}$, and $\theta_{q}\left(\beta\right)=\arctan\left[J_{0}(\beta) \tan \theta \right]$, with $\tan (\theta) = q_y/q_x$. These spinors
will be useful later on to compute the optical conductivity.

It is very relevant to note here that the pristine gap of hBN, the one
mentioned in the Introduction, is actually a sum of two contributions:
the DFT band gap $E_g$, and a correction coming from the interaction
between electrons~\cite{Ventura_2019}. The latter, that we will call
exchange correction and denote by $\Delta_{\text{ex}}$, is computed
in Section~2 of the Supporting Information (SI). The true gap of hBN is
then $\Delta = E_g + \Delta_{\text{ex}}$, and in the presence of the
potential, both contributions renormalize with $\beta$.

\subsection{Excitonic binding energies}

\begin{figure*}[t]
\centering
\includegraphics[width=0.80\textwidth]{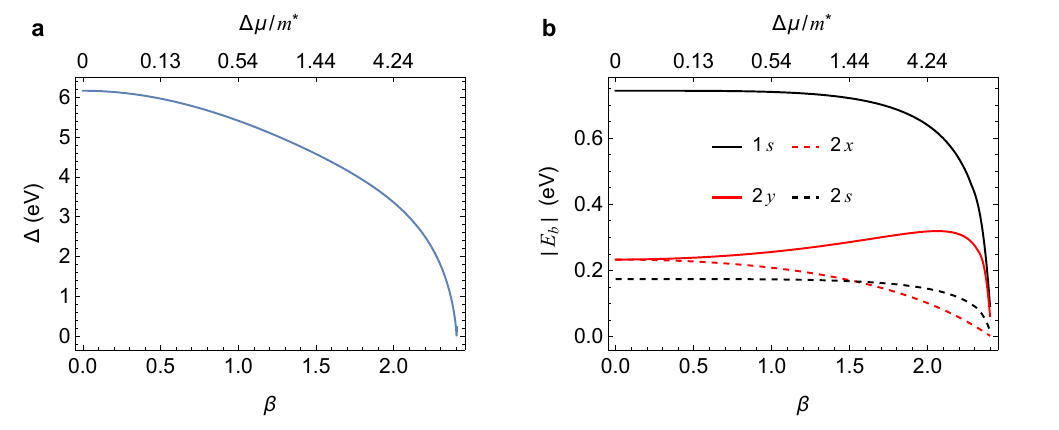}
\caption{(a) Total band gap of an hBN layer subjected
to an 1D periodic potential.
(b) Binding energies of the excitonic states $1s$ (black continuous line),
$2x$ (red dashed line), $2y$ (red continuous line) and $2s$ (black dashed
line) as a function of $\beta$.
In both panels, the top horizontal axis displays the value of the effective
mass anisotropy defined as $\Delta \mu = (m_y-m_x)/m^*$. The parameters
used for hBN are~\cite{Ferreira_2019}: $E_g=3.92$\,eV, $\hbar v_F = 5.06$\,eV$\cdot$\AA, and $r_0=10$\,{\AA}. For the surrounding medium we assume air with
$\varepsilon_1 = 1$, while for the quartz substrate $\varepsilon_2 (\omega \to 0) = 3.9$.}
\label{fig:Eb_of_beta}
\end{figure*}

The external 1D potential influences the electronic properties of hBN, as
we discussed in the previous section, in a way that depends
only on the parameter $\beta$, that in turn depends only on the product
between the period and strength of the potential, and not on both separately.
Hence, for a complete analysis, it suffices to study the system varying 
$\beta$. As $J_{0}(\beta) \to 0$, the dispersion becomes more and more
anisotropic, and the DFT gap closes. To study the effects of the potential
in the electronic and excitonic properties of the system
at hand, we need to compute the
\emph{total} band gap $\Delta$,
and the binding energies of the excitonic states $1s$, $2x$, $2y$ and $2s$
for varying $\beta$, from the case $\beta=0$ (without potential), to
$\beta=2.3$ (sufficiently close to the first zero of the Bessel function
$J_0(\beta)$). 
To describe the excitonic physics, we adopted the approach
of the Wannier equation $E_{b} \psi = H \psi$, where $H$ is the Hamiltonian
for the electron--hole pair, $\psi$ is the exciton wavefunction and $E_b$
its binding energy, and the electrostatic attraction is given by the
Rytova--Keldysh potential $V_{\text{RK}}(r) = 
-\frac{e^{2}}{4 \pi \varepsilon_0 r_{0}}
\frac{\pi}{2}
\left[\mathbf{H}_{0} \left(\frac{\kappa r}{r_{0}}\right) -
Y_{0} \left(\frac{\kappa r}{r_{0}}\right)\right]$,
where $\kappa$ is the arithmetic average between the static permittivities
of the substrate and the surrounding medium, $\mathbf{H}_{0}$ is the Struve
function of zeroth order, $Y_{0}$ is the Bessel function of the second kind
and zeroth order, $r_0$ is a screening parameter of hBN, $e$ is the elementary
charge, and $\varepsilon_{0}$ the vacuum permittivity. To solve the Wannier
equation, we apply the variational method for the first four excitonic states.
For more details, we refer to the Methods section.
The results are displayed in Figure~\ref{fig:Eb_of_beta}.

It can be seen in Figure~\ref{fig:Eb_of_beta}a that by manipulating the
value of $\beta$, we can tune the \emph{total} band gap.
When $\beta$ reaches the zero of $J_0(\beta)$, the gap closes. Regarding
the binding energies, Figure~\ref{fig:Eb_of_beta}b shows that the anisotropy
lifts the degeneracy between the states $2x$ and $2y$. Also, the binding
energies of the states $1s$, $2x$ and $2s$ decrease as we increase $\beta$,
but that of the $2y$ state is non-monotonic. Nonetheless, the state $2y$ is
always the first excited state,
as it is always the one with second highest binding energy
(in absolute value).
Initially, the second excited state is $2x$, but for $\beta \approx 1.5$,
the $2s$ takes its place.

\subsection{Optical conductivity and Absorption}\label{subsec:conductivity_and_absorption}

\begin{figure*}[t]
\centering
\includegraphics[width=1.0\textwidth]{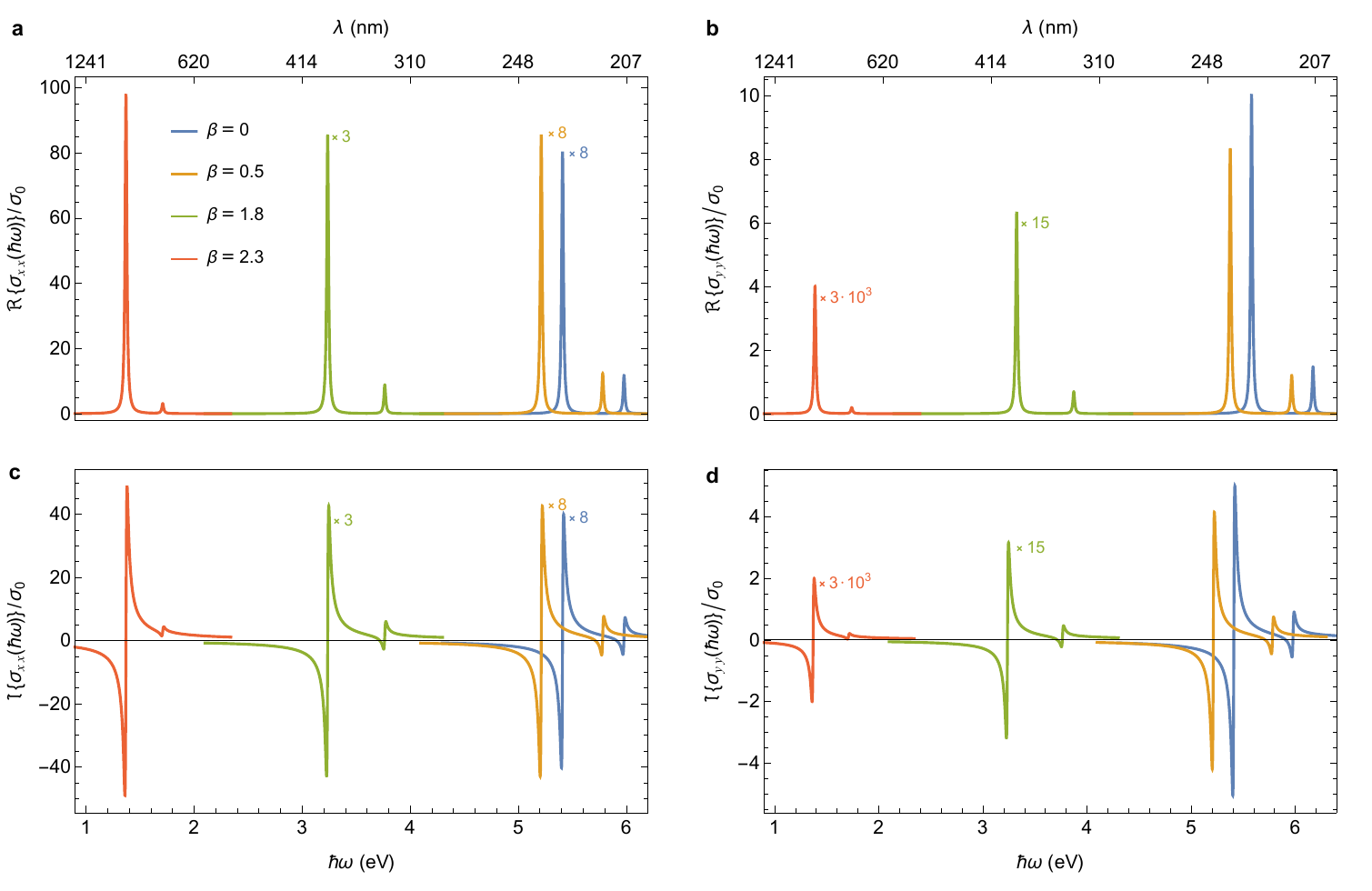}
\caption{Sub-gap optical conductivity of hBN under a 1D periodic potential,
as depicted in Figure~\ref{fig:hBN_monolayer}. Panels~(a) and (b) show the
real and imaginary parts of the conductivity, respectively, for light
polarized along the $x$-axis. Panels~(c) and (d) show the real and imaginary
parts of the conductivity, respectively, for light polarized along the
$y$-axis. Results for different $\beta$ have been rescaled as noted in each
panel, for clarity. In eq~\ref{eq:optical_conductivity} we use
$\Gamma=10$\,meV.}
\label{fig:conductivity}
\end{figure*}

A recent experimental work~\cite{Elias_2019} showed that single-layer hBN
displays optical response in the UV range. The measured photoluminescence
exhibits two sharp peaks at a frequency around $6.1$\,eV. These peaks are
excitonic in nature, and their theoretical prediction agrees very well with
the experimental measurement~\citen{Ventura_2019}. Here we apply the same
theory as Ventura et al.~\citen{Ventura_2019} to compute the optical 
absorption of monolayer hBN subjected to the external periodic potential.
But first, we computed the optical conductivity using
\cite{Pedersen_2015}
\begin{equation}\label{eq:optical_conductivity}
\sigma (\omega) = 
\frac{e^2}{8 \pi^3 i \hbar} 
\sum_\nu \left[ 
\frac{ E_\nu|\Omega_\nu|^2 }{E_\nu - (\hbar \omega+i\Gamma)} + 
\frac{ E_\nu|\Omega_\nu|^2 }{E_\nu +  \hbar \omega + i\Gamma} 
\right],
\end{equation}
where $E_\nu$ is the energy of the excitonic state $\nu$,
and $\Omega_\nu$ is a quantity related to the Berry connection, computed
from the excitonic wavefunction and the dipole matrix element, whose
calculation is explained in the Methods section, with further details in
Section~S3 of the SI. The conductivity computed through
eq~\ref{eq:optical_conductivity} for selected values of $\beta$, is
displayed in Figure~\ref{fig:conductivity}. Since the optical conductivity
is complex, we plot separately the real part in Figure~\ref{fig:conductivity}a
and Figure~\ref{fig:conductivity}c, and the imaginary part in Figure~\ref{fig:conductivity}b and Figure~\ref{fig:conductivity}d. In the
calculation we focus on the excitonic states $1s$ and $2s$, while the states
$2x$ and $2y$ do not contribute to the conductivity,
as explained in Section~3.3 of the SI.
We also considered both cases of light polarized along the $x$ and $y$ axes.
The results we obtained for the conductivity without 
potential, corresponding to the blue lines in each panel of 
Figure~\ref{fig:conductivity}, agree qualitatively with those obtained by
Ferreira et al.~\citenum{Ferreira_2019} via the Bethe--Salpeter equation
method. To be more precise, the behavior of the real part of the conductivity
that consists of two peaks centered at the excitonic energy levels in
Ferreira et al.~\citenum{Ferreira_2019} is similar to what we compute,
and also the imaginary part that changes sign at those levels is identical
to what we see in this work. We observe in Figure~\ref{fig:conductivity}
a tendency of a redshift in the peaks as we increase $\beta$, and consequently
the anisotropy of the system. We note that the energy level $E_{\nu}$ of the
excitonic state $\nu$ is given by $E_{\nu} = E_{b,\nu} + \Delta$, where
$E_{b,\nu}$ is the respective binding energy and $\Delta$ the
\emph{total} band gap, quantities computed previously.
For light polarized along the $x$-axis, we observe in 
Figure~\ref{fig:conductivity}a and in Figure~\ref{fig:conductivity}c that
the peaks of the conductivity coming from the $1s$ and $2s$ states are 
enhanced as we increase $\beta$. On the other hand, the peaks in the
conductivity for light polarized along the $y$-direction are attenuated.

One can notice also in Figure~\ref{fig:conductivity}b that for the case
$\beta=2.3$, the excitonic states $1s$ and $2s$ become so close, that the
imaginary part of the conductivity $\Im \{ \sigma(\omega) \}$ has only one
zero. We expect that this influences the character of the polaritons, since
the sign of $\Im \{ \sigma(\omega) \}$ dictates which kind of polaritons,
transverse magnetic (TM) or transverse electric (TE), the 2D material
supports, as we will explain next.

\begin{figure}[h!]
\centering
\includegraphics[width=0.5\textwidth]{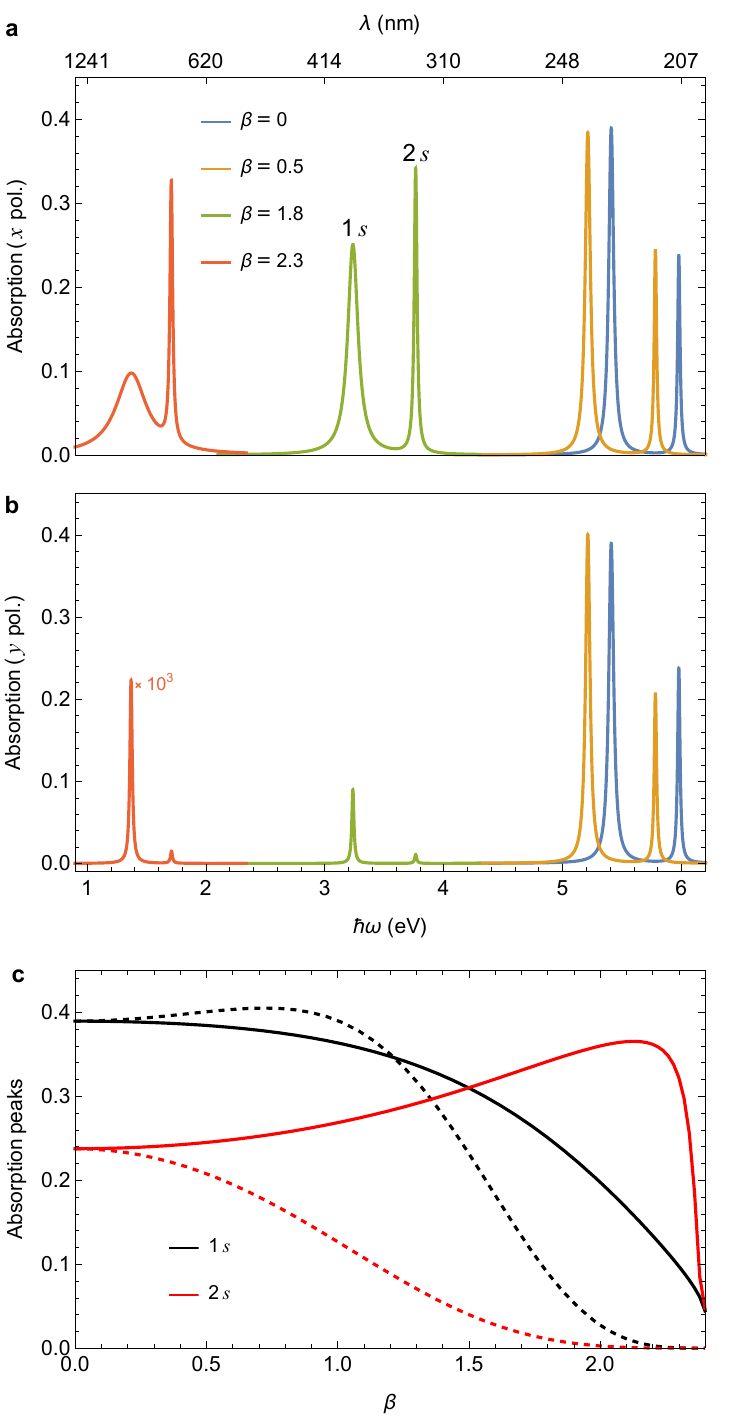}
\caption{Sub-gap optical absorption of hBN under a 1D periodic potential.
Panels~(a) and (b) show the absorption for light polarized along the $x$
and $y$-directions, respectively. Each color corresponds to a different
value of $\beta$, where $\beta=0$ corresponds to the case with no potential.
We use the same values for $\beta$ as in Figure~\ref{fig:conductivity}.
For each realization, two peaks are observed. The one occurring at lower
energy corresponds to the $1s$ state, while the other corresponds to the
$2s$ state. Panel~(c) shows the values of the absorption peaks due to the
excitonic states $1s$ (black lines) and $2s$ (red lines) as a function of
$\beta$ for light polarized along the $x$ (continuous lines) and $y$
(dashed lines) directions, respectively. Note that for each value of
$\beta$ the peaks occur at different energies, as one can see in
panels~(a) and (b). 
For the quartz substrate we use $\varepsilon_2 (\omega)
\approx 2.16$, an estimated average of the permittivity over a wide range
of optical frequencies.}
\label{fig:absorption}
\end{figure}

We can now compute the optical absorption at normal incidence using
Fresnel's coefficients and the fact that the absorption,
the reflection and the transmission add up to one, as in
eq~\ref{eq:absorption}. Figure~\ref{fig:absorption}a and
Figure~\ref{fig:absorption}b display the optical absorption for the same
values of $\beta$ as in Figure~\ref{fig:conductivity}, for light polarized
along the $x$ and $y$-directions, respectively. Interestingly, the behavior
of the absorption is not as simple as that of the conductivity. For instance,
let us look at Figure~\ref{fig:absorption}a. The peak due to the $1s$ state
seems to attenuate from right to left, as we increase $\beta$. In contrast,
the peak due to the $2s$ state seems to intensify. To study this behavior
more carefully, in Figure~\ref{fig:absorption}c we plot the value of the
absorption peaks due to the $1s$ (black lines) and $2s$ (red lines) states
separately, for polarization along the $x$ (continuous lines) and $y$
directions (dashed lines), as a function of $\beta$, which is allowed to
vary continuously. Indeed, we see that for polarization along the $x$ axis
the $1s$ peak decreases monotonically as we approach the zero of $J_0(\beta)$,
while the $2s$ peak tends to increase, until it reaches a maximum right
before the zero of the Bessel function. Interestingly, we can have a $2s$
peak more intense than a $1s$ peak. This phenomenon has already been
observed experimentally~\cite{Ju_2017} and explained 
theoretically~\cite{Henriques-PRB2022}.
This can be physically clarified by the fact that for high
anisotropy in the dispersion and $x$ polarization, the oscillator strength
at the $1s$ resonance is very high, thus the material behaves just like a
surface of a very high conducting material, such as a metal, reflecting
most part of the incident radiation. This can be seen by close inspection
of eq~\ref{eq:absorption}, which we do at the end of Section~S3 in the SI.
One can also observe that the absorption for light polarized along the
$y$-direction is not as intense as for light polarized along the
$x$-direction. Furthermore, from Figure~\ref{fig:absorption}c, for
light polarized along the $y$-direction we can note that it is now
the $2s$ peak that decreases monotonically, while the $1s$ peak is
maximum for $\beta \approx 0.9$.

\subsection{Exciton polaritons}\label{subsec:exciton_polaritons}

Polaritons are light--matter hybrids, arising from the strong-coupling of
elementary excitations of a 2D material with light. Polaritonic modes are
characterized by an evanescent decay away from the material and a strong
enhancement of the EM field in its vicinity. There are many kinds of
polaritons, like phonon polaritons, plasmon polaritons, magnon polaritons
and exciton polaritons~\cite{basov_nanoph10}. Here we are concerned only
with exciton polaritons. The dispersion relation of the polariton depends
on the polarization of the EM field. For TE modes, the electric field is
in the plane of the 2D material, and perpendicular to the wavevector of
the polariton, while the magnetic field lies on the $xz$-plane. The
dispersion relation for TE modes is 
\begin{equation}\label{eq:TE_modes_disp_rel}
\kappa_1 + \kappa_2 - i \omega \mu_0 \sigma_{yy}(\omega) = 0, 
\end{equation}
where $\varepsilon_j$, $j=1,2$, is the relative electric permittivity of
the two media cladding the hBN layer in the frequency region herein
considered, $\kappa_j = \sqrt{q^2 - \varepsilon_j \omega^2/c^2}$, $\omega$
is the angular frequency of the polariton, $q$ is the in-plane wavenumber,
and $\mu_0$ is the vacuum permeability.

For TM modes, it is the magnetic field that lies on the plane of the 2D
material, while the electric field is in the $xz$-plane. Their dispersion
relation is
\begin{equation}\label{eq:TM_modes_disp_rel}
\frac{\varepsilon_1}{\kappa_1} + 
\frac{\varepsilon_2}{\kappa_2} + 
i \frac{\sigma_{xx} (\omega)}{\varepsilon_0 \omega} = 0.
\end{equation}
As a rule of thumb, TE modes are supported when the imaginary part of the
conductivity is negative, whereas TM modes are supported when the imaginary
part of the conductivity is positive. The solutions of the previous two
equations, that can only be obtained numerically, gives the dispersion
relation $\omega(q)$, where the wavenumber has now an imaginary part,
since the conductivity is complex. It is important to highlight that in
eqs~\ref{eq:TE_modes_disp_rel} and \ref{eq:TM_modes_disp_rel} two different
conductivities appear. In isotropic systems, such as the case of monolayer
hBN, the conductivities $\sigma_{xx}$ and $\sigma_{yy}$ coincide. However,
they are in general different, as we have seen in the previous section.

Regarding the TE modes, we observe that they are either very poorly
confined and present a similar dispersion to free light as
expected~\cite{Bludov-primer,GoncalvesPeres:2016}, or they do not exist at
all (that is, eq~\ref{eq:TE_modes_disp_rel} does not present any solution).
Hence, we omit the analysis for TE modes, as they are not relevant. We
focus then on TM modes solely. To obtain the dispersion relation
$\omega(q)$ for TM polarized polaritons, we solve numerically
eq~\ref{eq:TM_modes_disp_rel} in the regions where $\Im \{\sigma(\omega)\}
> 0$, for different values of $\beta$. In general, the imaginary part of
the conductivity admits more than one zero, and in our case, we have
observed in Figure~\ref{fig:conductivity} that it has either two or one.
The main region of frequencies of interest is that between the first and
second zeros, in the case $\Im \{\sigma(\omega)\}$ does have two zeros.
After the third zero, we are very near the bottom of the conduction band,
where the sub-gap conductivity we computed herein is not accurate. To
compute the conductivity above the gap, one needs to also consider
unbound states~\cite{Ivchenko_2021}.

\begin{figure*}[t]
\centering
\includegraphics[width=1.0\textwidth]{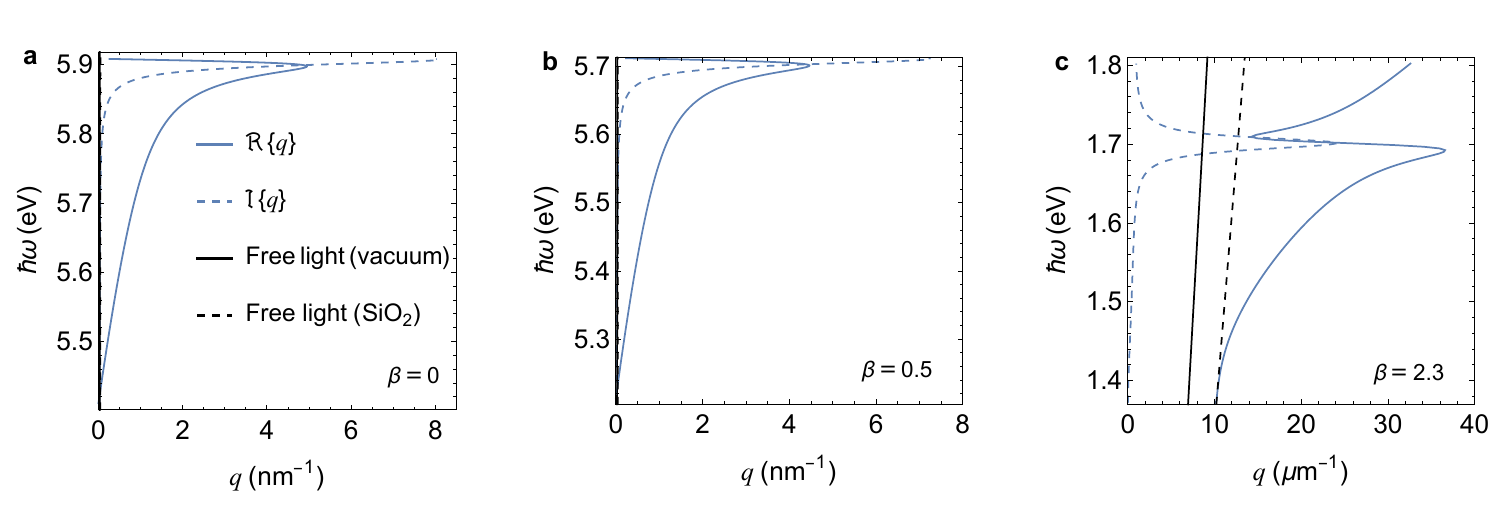}
\caption{Dispersion relation $\omega(q)$ for TM exciton polaritons, obtained
by solving numerically eq~\ref{eq:TM_modes_disp_rel} for different values of
$\beta$. The solution is the frequency as a function of the wavenumber, that
has a real part (solid blue line) and an imaginary part (dashed blue line).
For comparison, we show the dispersion relation for free propagating light
in vacuum (solid black line) and in quartz (dashed black line). The case
$\beta = 0$ represents monolayer hBN without potential on top.
For the quartz substrate, we use $\varepsilon_2 (\omega)
\approx 2.16$.}
\label{fig:polaritons_disp_rel}
\end{figure*}

The dispersion relation of TM modes is presented in 
Figure~\ref{fig:polaritons_disp_rel}, for three different values of
$\beta$. The case $\beta=0$ (Figure~\ref{fig:polaritons_disp_rel}a)
represents hBN with no potential on top, and we also choose an intermediate
case $\beta = 0.5$ (Figure~\ref{fig:polaritons_disp_rel}b), and an extreme
case $\beta=2.3$ (Figure~\ref{fig:polaritons_disp_rel}c). The case without
potential and the intermediate case share similar features, namely a
dispersive behavior very distinct from free light, with the real part of
the wavenumber $\Re\{q\}$ reaching a high maximum value at a specific
frequency, and the high degree of confinement given the high values that
$\Im\{q\}$ can reach. However, the $\beta = 2.3$ case of
Figure~\ref{fig:polaritons_disp_rel}c is qualitatively different. This is
due to the fact that in this case the imaginary part of the conductivity
has only one zero and the conductivity peaks at higher values. As a 
consequence, the dispersion behaves like free light in quartz for low
frequencies and then behaves like the previous two cases, with $\Re\{q\}$
achieving a maximum around $36 $~$\mu$m$^{-1}$. Then, the dispersion bends
until $\Re\{q\}$ reaches a minimum of 14\,$\mu$m$^{-1}$, and afterwards it
increases back again. 
This dispersion back-bending is due to the $2s$ resonance,
where the oscillator strength increases. The connection between these two
effects has been predicted before~\cite{Wolff:2018}. The degree of
confinement is also lower in this case, as can be seen by the different
wavenumber scales (nm$^{-1}$ in Figure~\ref{fig:polaritons_disp_rel}a and
Figure~\ref{fig:polaritons_disp_rel}b, $\mu$m$^{-1}$ in 
Figure~\ref{fig:polaritons_disp_rel}c). To gain intuition on the
confinement of the polariton, we show in Figure~S2 a spatial map of the
EM field of a TM mode in the vicinity of the interface between air and
the substrate. 

\section{Conclusions}\label{sec:conclusion}

With the advent of the experimental realization of 2D materials, the way
of manipulating the electronic properties of 2D materials by applying an
additional periodic modulation has gained significant attention.
Notwithstanding the current popularity of this topic, the idea of forming
superlattices to manipulate the properties of materials dates back to the
end of the last century, with studies on plasmonic excitations in
externally modulated 2D electron-gases~\cite{Zyl_1999,Fessatidis_1991}.
The periodic modulation can result from combining two monolayers to
form a bilayer structure and applying a twist to form a moir{\'e}
superlattice~\cite{Chu_2020,Yan_2019}, or even by combining bulk systems,
like for example two hBN slabs, bringing their surfaces together and
applying a twist~\cite{Lee_2021}. 

In our case, the external potential can be imposed by a periodic patterned
back/top gate,
as sketched in Figure~\ref{fig:hBN_monolayer}a.
The tuning of the potential has a high degree of flexibility, since our
predictions depend only on the product of the strength of the potential
and its period, but not on both independently. In a typical
3D semiconductor, such as \ch{GaAs} or \ch{CdTe},
the exciton size is of the order of 
5--12\,nm.
For the period, we need to choose a high enough value compared to the
exciton radius, so of the order of $40$\,nm. For this length scale,
and to have $\beta\sim1$, we need a potential amplitude of the order
of $40$\,meV, which is feasible in experiments.
With these values for the strength of the potential,
an hBN flake will not experience atomic reconstruction, as the
boron-nitrogen covalent bond in hBN is very strong.
However, our theory does not make predictions when $\beta$ matches
exactly the zero of the Bessel function, since the renormalized effective
mass along $x$ vanishes and that along $y$ diverges, rendering the
Wannier equation description inapplicable. Also, the simplicity of the
potential allowed for an analytical expression for the renormalized
low-energy Hamiltonian, but it comes with the disadvantage that a cosine
potential is more difficult to implement experimentally. We remark that
our theory also applies to a square periodic potential, more common in
experiments. Hence, our results widely apply to a variety of periodic
potentials
Our semi-analytic treatment can also be extended to
other 2D semiconductors and substrate materials. 

This work presents a way of changing the frequency of the excitonic
resonances of hBN, as an alternative to the usual one of changing the
substrate. The redshift in the resonances as we increase the strength
(or period) of the potential makes it possible to make hBN optically
non-trivial in between the near- and the deep-UV. Furthermore, the
system response is different depending on the polarization of the
incident field, which is the essential characteristic of a grid wire
polarizer.
Hence, our setup is a possible prototype for the
fabrication of very efficient grid polarizers. Beyond that, the
characteristics of tunable exciton-polaritons, namely the
ultra-confinement of the EM field and the tuning of the frequency
window within which they can be excited, renders our setup also a
prototype for the fabrication of sensitive biosensors for organic
molecules with different resonant spectra.

The study of excitonic and optical properties in twisted bilayer systems
is a natural extension of this work. A key feature in these systems is the
emergence of flat electronic bands, enhancing interactions and promoting
many-body correlated states, which offer a plethora of exotic phenomena in
physics~\cite{Jarillo_2018,Cao_2018}. There are already studies of excitons
in these systems~\cite{Malic_2023,Roman-Taboada_2023} that predict that the
optical spectrum of the bilayer system is very different from that of the
isolated monolayer. Furthermore, plasmon polaritons have been studied in
twisted bilayer graphene as well~\cite{Stauber_2016}, where the appearance
of electronic flat bands leads to plasmon polaritons with almost constant
dispersion. However, polaritonics in the UV range within the realm of
twisted bilayer systems remains largely unexplored. With the experimental
and technological tools available nowadays, one can expect solid developments
in the field of 2D moir{\'e} structures in general, and in particular in the
study of excitons~\cite{Wu_2022,Zhao_2023}. 

\section{Methods}
\label{sec:methods}

\subsection{Renormalized Hamiltonian}\label{sec:renormalized_hamiltonian}

The electronic properties of independent electrons in 2D materials 
can be described, approximately, in terms of tight-binding
models; for the tight-binding description to be as realistic as possible,
its parameters are determined by first principal methods, such as DFT and
$GW$.
In general, these models use a fairly large number of bands, which prevents
analytical calculations of the materials' properties. Fortunately, the
physics of excitons is often determined by effective models with only two
bands, whose Hamiltonian can be written using Pauli matrices. These are
the cases of graphene and hBN. The difference between the two systems lies
in the presence of a band gap in the latter. With all this in mind, an
effective, low-energy, independent-electron Hamiltonian can be written
as~\cite{CastroNeto-RMP}
\begin{equation}
H_{0} = -i\hbar v_{F}
\left(\sigma_{x} \partial_x +\sigma_{y} \partial_y\right) + 
M\sigma_{z} \,,
\end{equation}
where $E_g=2M$ is the band gap obtained within DFT~\cite{Ferreira_2019}, and the $\sigma$'s are the Pauli matrices.
The low-energy dispersion relation reads
\begin{equation}\label{eq:DR-isotropic}
    E_{s} (\mathbf{q})= s \frac{1}{2} \sqrt{E_g^2 + 4 \hbar^2 v_F^2 (q_x^2 + q_y^2)} \approx s \left(\frac{E_g}{2} + \frac{\hbar^2 |\mathbf{q}|^2}{2 m^*}\right)\,,
\end{equation}
where the wavevector $\mathbf{q}$ is measured from the $\mathbf{K}$ point
in the BZ (see Figure~\ref{fig:hBN_monolayer}b), while
$m^*=E_g/2v_F^2$ is the common effective mass of electrons in the conduction
band ($s=+1$) and holes in the valence band ($s=-1$), respectively. The
approximation of parabolic bands is valid when the DFT band gap is large
and when looking at electronic states with small momentum (measured from
the point around which we are expanding the Hamiltonian).

Since we want to describe the effect of an external potential on the
electronic properties of electrons in hBN (including its excitonic 
properties), we introduce an external electrostatic potential $V(x)$,
so that the total Hamiltonian reads
\begin{equation}
    H = H_0 + V(x).
\label{eq:fullH}
\end{equation}
The potential varies only along the $x$-direction and has period $L$,
that is, $V(L+x)=V(x)$. This external potential then creates a
superlattice, leading to a mini-BZ of size $[-\pi/L,\pi/L[$, as in
Figure~\ref{fig:hBN_monolayer}c, and the appearance of a set of
mini-bands. This problem is similar to graphene under an external
potential~\cite{Louie_2008}, but with an extra term that confers a
mass to the electrons thus opening an energy gap at $K$ of the original
BZ. By introducing the external potential $V(x)$, we add an extra
periodicity in the system in addition to the original lattice periodicity
(already encoded in the original low-energy Hamiltonian). As noted above,
we form a 1D superlattice whose reciprocal lattice vectors are given by
$ \mathbf{G}_{\ell} = \ell G_0 \hat{x}$, with $G_0=2\pi/L$ and $\ell$
an integer. 

It would be convenient if we could remove the potential from the 
Hamiltonian eq~\ref{eq:fullH}, while maintaining the Dirac-like form of
$H_0$. This is possible via a unitary transformation written in terms of
$V(x)$ and given by~\cite{Louie_2008}
\begin{equation}
U_{\alpha}=\frac{1}{\sqrt{2}}\left(\begin{array}{cc}
\mathrm{e}^{-i\alpha(x)/2} & -\mathrm{e}^{i\alpha(x)/2}\\
\mathrm{e}^{-i\alpha(x)/2} & \mathrm{e}^{i\alpha(x)/2}
\end{array}\right),
\end{equation}
where $\alpha(x) = (2/\hbar v_{F}) \int_{0}^{x} \mathrm{d}x^{\prime}
V(x^{\prime})$.
Now we rotate the total Hamiltonian in eq~\ref{eq:fullH}
to $H \to H'= U_{\alpha}^{\dagger} [H_{0}+V(x)]U_{\alpha} =
U_{\alpha}^{\dagger} H_{0} U_{\alpha}+V(x)$. The first term is a sum of
three contributions, each one going with a Pauli matrix, and it needs to
be worked out contribution by contribution. The contribution that goes
with $\sigma_z$ is the simplest, and reads

\begin{equation}\label{eq:rotated_Hz}
MU_{\alpha}^{\dagger}\sigma_{z}U_{\alpha}=
\left(\begin{array}{cc}
0 & -M \mathrm{e}^{i\alpha}\\
-M \mathrm{e}^{-i\alpha} & 0
\end{array}\right).
\end{equation}
To work out the contribution that goes with $\sigma_y$, we
compute the matrix element of that term between two general functions
$f$ and $g$ and do an integration by parts in $y$ to write

\begin{equation}\label{eq:rotated_Hy}
H_{y} \left[f,g\right] = 
\frac{\hbar v_{F}}{2} \int \mathrm{d}\mathbf{r} f 
\left(\begin{array}{cc}
0 & -\mathrm{e}^{i\alpha(x)}\\
\mathrm{e}^{-i\alpha(x)} & 0
\end{array}
\right) 
\frac{\partial g}{\partial y} -\frac{\hbar v_{F}}{2}
\int \mathrm{d}\mathbf{r} \frac{\partial f}{\partial y}
\left(\begin{array}{cc}
0 & -\mathrm{e}^{i\alpha(x)}\\
\mathrm{e}^{-i\alpha(x)} & 0
\end{array}\right)g
\end{equation}

To compute the contribution with $\sigma_x$, we need to take
more care, as the unitary transformation depends on $x$. In this way, when
we rotate that term and integrate by parts, there appear terms where the
derivative $\partial_x$ acts only on $U_{\alpha}$. When those terms are
evaluated, they sum to $-V(x)$, which cancels the potential term in the
total Hamiltonian. Then, the terms with $\sigma_x$ that survive read

\begin{equation}\label{eq:rotated_Hx}
H_{x}\left[f,g\right ] 
=\frac{-i\hbar v_{F}}{2}
\int \mathrm{d}\mathbf{r} f 
\left(\begin{array}{cc}
0 & \mathrm{e}^{i\alpha(x)}\\
\mathrm{e}^{-i\alpha(x)} & 0
\end{array}
\right)
\frac{\partial g}{\partial x} +
\frac{i \hbar v_{F}}{2}
\int \mathrm{d}\mathbf{r}\frac{\partial f}{\partial x}
\left(\begin{array}{cc}
0 & \mathrm{e}^{i\alpha(x)}\\
\mathrm{e}^{-i\alpha(x)} & 0
\end{array}\right)
g
\end{equation}

Focusing on electronic states whose wavevector $\mathbf{k} = \mathbf{q} +
\ell \mathbf{G}_0/2$, with $|\mathbf{q}| \ll G_0$, we obtain an analytical
expression for the renormalized low-energy Hamiltonian, that is, the
initial low-energy Hamiltonian but with renormalized parameters. For
that, we use the following states as the basis functions
\begin{equation}
\phi_1=
\left(\begin{smallmatrix}
1 \\\\
0
\end{smallmatrix}\right) 
\mathrm{e}^{i (\mathbf{q} + \ell \mathbf{G}/2) \cdot \mathbf{r}} 
\quad \text{ and } \quad
\phi_2=\left(\begin{smallmatrix}
0 \\ \\
1
\end{smallmatrix} \right)
\mathrm{e}^{i (\mathbf{q} - \ell \mathbf{G}/2) \cdot \mathbf{r}}
,
\end{equation}
and expand the exponential in a Fourier series as
\begin{equation}\label{eq:beta}
\mathrm{e}^{i \alpha(x)} = 
\sum_{\ell=-\infty}^{\infty}
J_{\ell}\left(\beta\right)
\mathrm{e}^{i\ell G_{0}x}, 
\quad 
\beta \equiv \frac{V_{0}L}{\pi\hbar v_{F}},
\end{equation}
where $J_\ell$ is the Bessel function of the first kind of order $\ell$.
The previous identity holds only for a cosine potential.
After applying the unitary transformation, we compute the matrix elements
for the previous basis states
for all the three terms, by replacing $f$ and $g$ with the
basis functions. For that, one also has to compute the matrix elements
$\langle \phi_i |\mathrm{e}^{\pm i \alpha(x)}| \phi_j \rangle = J_0(\beta)
\delta_{ij} + J_{\pm \ell}(\beta) \delta_{i1} \delta_{j2} +
J_{\mp \ell}(\beta) \delta_{i2} \delta_{j1}$, and replace the spatial 
derivatives as $\partial_x \to i\left(q_{x} \pm \ell G_{0}/2\right)$ and
$\partial_y \to i q_y$. After computing all the matrix elements and summing
all the contributions, one obtains for the Hamiltonian
\begin{equation}
\bar{H} = 
\left(\begin{array}{cc}
\hbar v_{F}\left(q_{x}+\ell\frac{G_{0}}{2}\right) & 
i\hbar v_{F}q_{y}J_{\ell}(\beta)-MJ_{\ell}(\beta)\\
-i\hbar v_{F}q_{y}J_{\ell}(\beta)-MJ_{\ell}(\beta) & 
\hbar v_{F}\left(-q_{x}+\ell\frac{G_{0}}{2}\right)
\end{array}\right),
\end{equation}
and changing to a basis made of a symmetric and anti-symmetric combination
of the previous states, we obtain the renormalized low-energy Hamiltonian
\begin{equation}
\label{eq:renormalized_H_all_bands}
\bar{H}_{\ell} =
\frac{\ell}{2}G_{0}\hbar v_{F} +
\hbar v_{F}[\sigma_{x}q_{x}+\sigma_{y}J_{\ell}(\beta)q_{y}] -
\sigma_{z}MJ_{\ell}(\beta).
\end{equation}
Here, the first term simply shifts the energies. Since $\ell$ is an integer,
we have now multiple bands, whose emergence can also be predicted by the
theory of nearly-free electrons. The comparison between the two treatments
is done in Section~1 of the SI.

To assess the validity of our results, we study eq~\ref{eq:renormalized_H_all_bands} in the limits $V_0 \to 0$ and
$L \to \infty$. In both limits, we should obtain the low-energy Hamiltonian
of hBN in the absence of the potential. Considering first the limit
$V_0 \to 0$ in eq~\ref{eq:renormalized_H_all_bands}, we obtain
\begin{equation}
\bar{H}_{\ell} =
\frac{\ell}{2}G_{0}\hbar v_{F}+\hbar v_{F}\sigma_{x}q_{x}
\end{equation}
for $\ell \neq 0$, and
\begin{equation}
\bar{H} =\hbar v_{F}[\sigma_{x}q_{x}+\sigma_{y}q_{y}]-\sigma_{z}M
\end{equation}
\noindent for $\ell = 0$. The latter case is simply the
original low-energy Hamiltonian of hBN in the absence of potential (with
a different sign for the mass term, that does not influence the physics
we study here). The former case does not reduce to zero, because it is
the Hamiltonian for electronic states with momentum connected to the $K$ point of the original hexagonal BZ by an integer multiple
of $G_0 \hat{x}/2$, hence the first term is just a global shift in
energies that comes from the shift in momentum and the second one is
an artifact of measuring momentum along the $x$ direction around the
edges of mini Brillouin zones of the superlattice that no longer exists

Considering now the case of $L \to \infty$, we should also
recover the original low-energy Hamiltonian, as this case corresponds to a
potential with infinite period, i.e., a constant potential. Looking again
at eq~\ref{eq:renormalized_H_all_bands}, we note that for each $\ell$ we
have a different Hamiltonian. Writing the total Hamiltonian including all
the bands as
\begin{align}
H_{\text{total}} &= 
\sum_{\ell} \bar{H}_{\ell} = 
\nonumber\\
&= \sum_{\ell} \frac{\ell}{2} G_{0}\hbar v_{F} +
\hbar v_{F} [\sigma_{x}q_{x}+\sigma_{y} q_{y} \sum_{\ell}J_{\ell}(\beta)] -
\sigma_{z} M \sum_{\ell} J_{\ell}(\beta) = 
\nonumber\\
& \overset{L\to \infty}{=} \hbar v_{F}
[\sigma_{x}q_{x}+\sigma_{y}q_{y}] - \sigma_{z}M,
\end{align}
\noindent where in the last equality we took the limit
$L \to \infty$, and used the fact that $G_0 \propto 1/L \to 0$ and the
property $\sum_{\ell} J_{\ell}(\beta) = 1$ of the Bessel functions of
the first kind. In this way, we recovered the original low-energy
Hamiltonian of hBN (again with the mass term sign flipped).
    
\subsection{Excitonic contributions to sub-gap optical response}\label{sec:sub_gap_optical_response}

The presence of the external potential $V(x)$ influences considerably the
electronic properties of hBN, by rendering the low-energy dispersion 
anisotropic. This brings consequences to the excitonic physics, and 
subsequently to the optical response. There are several approaches to the
calculation of the excitonic properties of a semiconductor or a band
insulator. Excitons with a radius large enough compared to the lattice 
spacing are denoted Wannier excitons. The general approach to determine
the exciton energy levels and wavefunctions is the solution of the
Bethe--Salpeter equation~\cite{Tenorio_Peres_2022} in momentum space.
This microscopic approach is powerful but often computationally demanding.
It would then be desirable to have in reach a simpler approach that still
captures the essence of the exciton physics. Starting from a microscopic
approach~\cite{Have_2019}, if 
the dispersion is quadratic in momentum and the DFT band
gap is large, then
the Fourier transform of the Bethe--Salpeter equation transforms into an
effective Schr\"{o}dinger equation with an electrostatic attractive
potential called the Wannier equation
(see further details in Section~2 of the SI).
The method of the Wannier equation has been benchmarked against the
Bethe--Salpeter equation with good results, as the latter equation
accounts for fine details only~\cite{Have_2019}.

The Wannier equation can be solved, among others, with the variational
method, which allows for a good approximation of the energies and
approximate analytical expressions for the excitonic wavefunctions.
This approach has previously been applied to black 
phosphorus~\cite{Mikhail_2021} and 
TMDs~\cite{Quintela_Peres_2020,Henriques_2021}, with significant
accuracy.

\subsubsection{Variational method}

The variational method~\cite{Sakurai} consists in using a trial wavefunction,
that depends on a set of parameters, to compute the expectation value of the
Hamiltonian $H$ and minimize it with respect to said parameters. 
Symbolically, we define a state $|\psi_{\alpha} (\mathbf{r}) \rangle$, with
$\alpha$ representing the set of unknown parameters. Then we compute 
$E_{\alpha} = \langle \psi_{\alpha} (\mathbf{r})|H|\psi_{\alpha} (\mathbf{r})\rangle$ and solve $\partial E/\partial \alpha = 0$ for
$\alpha$. Finally, we return to $E_{\alpha}$ and substitute the optimal
values for $\alpha$.

The Hamiltonian $H$ pertains to the system comprised of an electron--hole
pair, 
\begin{equation}
\label{eq:electron_hole_hamiltonian}
H=\frac{p_{x}^{2}}{2\mu_{x}}+\frac{p_{y}^{2}}{2\mu_{y}} + V_{\text{RK}}(r),
\end{equation}
where the first two terms correspond to the kinetic contribution, with
$\mu_{x}$ and $\mu_{y}$ being the electron--hole reduced masses along the
$x$ and $y$-directions, respectively, and the last term corresponding to
the electrostatic attraction between the electron and the hole in the 2D
material, here given by the electrostatic Rytova--Keldysh
potential~\cite{Tenorio_Peres_2022}
\begin{equation}
\label{eq:rytova_keldysh_potential}
V_{\text{RK}}(r)=-\frac{e^{2}}{4 \pi \varepsilon_0 r_{0}}\frac{\pi}{2}\left[\mathbf{H}_{0}\left(\frac{\kappa r}{r_{0}}\right)-Y_{0}\left(\frac{\kappa r}{r_{0}}\right)\right].
\end{equation}
In our work, the surrounding medium is air ($\varepsilon_1=1$), while for
the substrate we choose silicon dioxide (\ch{SiO2}). The static dielectric
function of quartz is $\varepsilon_2(\omega \to 0)=3.9$, and in the UV range
it is essentially constant, with $\varepsilon_2(\omega) \approx 2.16$. 
As mentioned previously in the "Excitonic binding energies" section, the
exciton dynamics is determined by the Wannier equation $E_{b} \psi = H \psi$,
which is a Schr{\"o}dinger-like equation, where $H$ is the Hamiltonian given
in eq~\ref{eq:electron_hole_hamiltonian}, $\psi$ is the exciton wavefunction
and $E_b = E - \Delta$ is the exciton binding energy, while $E$ is the energy
level of the exciton and $\Delta$ the total band gap. To obtain the energy
levels of the exciton we have to compute the binding energies and $\Delta$.
For the former, we apply the variational method to the first four excitonic
states $1s$, $2x$, $2y$ and $2s$, and use trial wavefunctions that resemble
the solutions of the 2D hydrogen atom, but taking into account the effect of
the anisotropy in the mass tensor. The form of each wavefunction is given in
Section~3.1 of the SI, eqs~S.10a-d, and the details of the calculations for
finding the variational parameters are given in Mikhail
et al.~\cite{Mikhail_2021}. The total band gap is the sum of the
renormalized DFT gap and the electron--electron exchange correction, both
contributions renormalizing with the potential (see Section~2 of the SI
for details).

\subsubsection{Optical conductivity}

Subjecting the excitonic material to an external and monochromatic EM field
with frequency $\omega$, the optical conductivity, within linear response
and in the dipole approximation, is given by
eq~\ref{eq:optical_conductivity}, and it can be obtained
within perturbation theory. However, the derivation is lengthy and not
straightforward, so for further details we refer to Pedersen
et al.~\cite{Pedersen_2015}.
The broadening parameter $\Gamma \ll \omega$ is the sum of the non-radiative
decay due to phonons, the non-radiative decay due to impurities and the
radiate decay, but its calculation is outside the scope of this work.
Therefore, we made an estimation based on previous works~\cite{Ventura_2019}.
Furthermore, $\Omega_\nu$ is related to the Berry curvature
$\boldsymbol{\Omega}_{cv,\mathbf{q}}= iV_{\text{UC}}^{-1} \int_{\text{UC}}
u_{+,\mathbf{q}}^*\nabla_{\mathbf{q}} u_{-,\mathbf{q}} \mathrm{d}\mathbf{r}$,
with $V_{\text{UC}}$ being the volume of the unit cell and
$u_{\pm,\mathbf{q}}$ the Bloch wavefunctions of the conduction and valence
bands, for $+$ and $-$, respectively, and the excitonic state $\nu$ through
\begin{equation}
\Omega_\nu = 
\sum_{\mathbf{q}} \Psi_{\nu} (\mathbf{q}) \hat{\mathbf{e}} \cdot 
\boldsymbol{\Omega}_{cv,\mathbf{q}},
\end{equation}
where $\Psi_{\nu} (\mathbf{q})$ is the wavefunction of the
excitonic state $\nu$ in momentum space, and the unit vector
$\hat{\mathbf{e}}$ is dictated by the polarization of the wave. We can
identify the Berry curvature $\boldsymbol{\Omega}_{cv,\mathbf{q}} $ with the
dipole matrix element between those bands. One simply considers $i$ times
the gradient operator in momentum space as the position operator
$\mathbf{r}$. In this way, $\boldsymbol{\Omega}_{cv,\mathbf{q}} =
\langle +, \mathbf{q} | \mathbf{r}| -, \mathbf{q} \rangle$, where the
bra/ket corresponds to the spinor of the Hamiltonian 
(eq~\ref{eq:renormalized_Hamiltonian_results}) for the conduction/valence
band.
Herein we consider light polarized along either the $x$ and the
$y$-directions. As the system is anisotropic, the conductivity will be
different depending on the polarization of the incident EM field. As
explained in Section~3 of the SI, we have two distinct conductivities
$\sigma_{xx}$ and $\sigma_{yy}$ for light polarized along the $x$ and
$y$-directions, respectively. Knowing the optical conductivity, we compute
the sub-gap optical absorption through
\begin{equation}
\label{eq:absorption}
\mathcal{A} = 1 - \mathcal{R} - \Re\left\{\sqrt{\frac{\varepsilon_2}{\varepsilon_1}}\right\}\mathcal{T},
\end{equation}
where the reflection and transmission coefficients $\mathcal{R}$ and
$\mathcal{T}$ follow from Fresnel's equations for normal incidence on
a conductive material cladded between two dielectrics~\cite{Bludov-primer}.
We note that the relative permittivities that enter in eq~\ref{eq:absorption}
are the dynamic dielectric functions of the respective materials. In the
electron--electron or electron--hole interaction, the relevant quantities are
the static permittivities of the materials. In the end, the dependence is
only on the average between the static permittivities, that we denoted by
$\kappa$ in eq~\ref{eq:rytova_keldysh_potential}.

\begin{acknowledgement}

N.~A.~M. is a VILLUM Investigator supported by VILLUM FONDEN (Grant No.~16498).
The Center for Polariton-driven Light--Matter Interactions (POLIMA) is funded by the Danish National Research Foundation (Project No.~DNRF165).
This project was supported by the Independent Research Fund Denmark (grant no. 2032-00045A).
N.~M.~R.~P. acknowledges support by the Portuguese Foundation for Science and Technology (FCT) in the framework of the Strategic Funding UIDB/04650/2020, COMPETE 2020, PORTUGAL 2020, FEDER, and through projects PTDC/FIS-MAC/2045/2021 and EXPL/FIS-MAC/0953/2021. N.~M.~R.~P. also acknowledges the Independent Research Fund Denmark (grant no. 2032-00045B) and the Danish National Research Foundation (Project No.~DNRF165).

\end{acknowledgement}

\begin{suppinfo}

Conciliating nearly-free electron theory with the analytical method, calculation of the electron--electron exchange correction to the electronic band gap, calculation of the optical conductivity, and spatial map of the electromagnetic field of the polariton.                                                                                                                                                                                                                                                                                                                                                                                             

\end{suppinfo}

\bibliography{achemso-demo}

\end{document}